\documentclass[draft,jgrga]{agu2001}
\usepackage [xdvi]{graphicx}

\authorrunninghead{RICCI, BRACKBILL, DAUGHTON, and LAPENTA}

\titlerunninghead{COLLISIONLESS RECONNECTION WITH GUIDE FIELD}

\journalid{}

\articleid{}{}

\paperid{}

\cpright{}{}

\received{} \revised{} \accepted{} \published{}

\authoraddr{P. Ricci,
Dipartimento di Energetica, Politecnico di Torino, C.so Duca degli
Abruzzi 24 - 10129 Torino, Italy.}

\authoraddr{J.U. Brackbill, W. Daughton, G. Lapenta,
Los Alamos National Laboratory, Los Alamos NM 85745 (jub@lanl.gov,
daughton@lanl.gov, lapenta@lanl.gov).}

\begin{document}

%
%

\title{Kinetic simulations of collisionless magnetic reconnection in presence of a guide field}


%
%


\author{Paolo Ricci $^{(1,2)}$, J.U. Brackbill $^{(3)}$,
W. Daughton $^{(3)}$, Giovanni Lapenta $^{(1,3)}$} \affil{1)
Istituto Nazionale per la Fisica della Materia (INFM) Politecnico
di Torino, Torino, Italy 2)  Dipartimento di Energetica,
Politecnico di Torino, Torino, Italy 3) Los Alamos National
Laboratory, Los Alamos, NM}

\begin{abstract}
The results of kinetic simulations of magnetic reconnection in
Harris current sheets are analyzed. A range of guide fields is
considered to study reconnection in plasmas characterized by
different $\beta$ values, $\beta > m_e/m_i$. Both an implicit
Particle-in-Cell (PIC) simulation method and a parallel explicit
PIC code are used. Simulations with mass ratios up to the physical
value are performed. The simulations show that the reconnection
rate decreases with the guide field and depends weakly on the mass
ratio. The off-diagonal components of the electron pressure tensor
break the frozen-in condition, even in low $\beta$ plasmas. In
high $\beta$ plasmas, evidence is presented that whistler waves
play a key role in the enhanced reconnection, while in low $\beta$
plasmas the kinetic Alfv\'en waves are important. The in-plane and
the out-of-plane ion and electron motion are also considered,
showing that they are influenced by the mass ratio and the plasma
$\beta$.

\end{abstract}

%
%

%

\begin{article}

\section{Introduction}

Magnetic reconnection causes global changes of the magnetic field
topology and of the plasma properties, and the conversion of
magnetic energy into plasma particle kinetic energy in form of
plasma jetting and plasma heating [{\it Biskamp}, 2000; {\it
Priest and Forbes}, 2000].

Magnetic reconnection is observed to occur in collisionless
plasmas over a wide range of $\beta$ values: In the geomagnetic
tail [{\it \O ieroset}, 2001], $\beta >> 1$; in the Earth's
magnetopause [{\it Nishida}, 1978], $\beta \approx 1$; in the
solar corona [{\it Priest}, 1982], in laboratory [{\it Gekelman et
al.}, 1991; {\it Yamada} 1999; {\it Egedal et al.}, 2001; {\it
Furno et al.}, 2003] and fusion plasmas [{\it Taylor}, 1986], in
extragalactic jets [{\it Romanova and Lovelace}, 1992; {\it
Blackman}, 1996], and in flares in active galactic nuclei [{\it
Lesch and Birk}, 1997], $\beta \leq 1$.

In the high $\beta$ case with zero guide field, the Geospace
Environment Magnetic (GEM) reconnection challenge [{\it Birn et
al.}, 2001; and references therein] has clarified the physics of
fast reconnection. The primary mechanism by which the frozen-in
condition is broken is given by the non-gyrotropic electron
pressure terms [e.g., {\it Hesse et al}, 1999;{\it Pritchett},
2001; {\it Ricci et al.}, 2002b]. The reconnection rate is then
enhanced thanks to the Hall term, which gives rise to the whistler
dynamics and decouples the electron and ion motion [e.g., {\it
Birn et al.}, 2001].

At lower $\beta$, the physics of fast reconnection plasmas is
still under investigation. Theoretical [e.g., {\it Pritchett},
2001] and experimental [{\it Yamada et al.}, 1997] results provide
strong evidence that fast reconnection still occurs in lower
$\beta$ plasmas, but the reconnection rate is reduced. However,
the scaling of the reconnection rate with the plasma $\beta$ and
with the mass ratio has not been clarified completely. Theoretical
studies [{\it Kleva et al.}, 1995; {\it Biskamp}, 1997; {\it
Rogers et al.}, 2001] have proposed Kinetic Alfv\'en Wave (KAW)
dynamics as the mechanism that enables fast reconnection in lower
$\beta$ plasmas, but the signature for this mechanism has been
observed only in fluid simulations [{\it Rogers et al.}, 2003].
Recently, for $\beta \approx 1$ plasmas, evidence has been
presented that the off-diagonal terms of the electron pressure
tensor break the frozen-in condition [{\it Hesse et al.}, 2002;
{\it Yin and Winske}, 2003] but it is not known what happens in
lower $\beta$ plasmas. The guide field allows drift motions that
are responsible for a typical asymmetry in the ion and electron
motion in the reconnection plane [{\it Hoshino and Nishida}, 1983;
{\it Hoshino}, 1987; {\it Pritchett}, 2001; {\it Yin and Winske},
2003] and the out-of-plane velocity is also affected by the plasma
$\beta$ [{\it Larrabee et al.}, 2003; {\it Nodes et al.}, 2003;
and references therein], but how the velocities depend on the mass
ratio and plasma $\beta$ has not been examined.

The aim of the present paper is to study magnetic reconnection in
plasmas with different $\beta$ values, using kinetic simulation to
study reconnection at low plasma $\beta$ with mass ratios up to
the physical value. The reconnection process is simulated using
two Particle-in-Cell (PIC) codes, which model both kinetic ions
and electrons. We use CELESTE3D, an implicit PIC code [{\it
Brackbill and Forslund}, 1985; {\it Vu and Brackbill}, 1992; {\it
Ricci et al.}, 2002a], which is particularly suitable for large
scale simulations with high mass ratios, and NPIC, a
two-dimensional massively parallel explicit code [{\it Morse and
Nielson}, 1971; {\it Forslund}, 1985], which is particularly
suitable for studies of microphysical processes on all scales. The
initial condition is a perturbed Harris sheet equilibrium and the
system is permitted to evolve freely. The plasma $\beta$ is
changed by varying the intensity of the initial guide field,
ranging from $\beta >> 1$ (no guide field case), to $\beta < 1$
(strong guide field).

It should be remarked that other physical systems have been
considered in the literature  in order to study reconnection in
low $\beta$ plasmas [{\it Bobrova et al.}, 2001;{\it Nishimura et
al.}, 2003; {\it Drake et al.}, 2003; {\it Rogers et al.}, 2003].
In Nishimura {\it et al.} [2003] a sheet pinch equilibrium is
considered, and the growth of the tearing and Buneman
instabilities are analyzed. In Drake {\it et al.} [2003] a double
current layer is studied with $\lambda=0.25 d_{i}$, with total
magnetic field $B$ and density constant. The three-dimensional
particle simulations performed therein show the development of
turbulence which collapses in structures where the electron
density is depleted. Rogers {\it et al.} [2003] consider fluid
simulations of current layers of width $\lambda= d_{i}$ and point
out that both the total $\beta$ and the $\beta$ based on the
reconnecting field ($\beta_x=8 \pi n_0 (T_i+T_e)/B_{x0}^2$) play
an important role in determining the structure of the out-of-plane
field and pressure profiles (in our simulation the total $\beta$
is varied, while $\beta_x$ is being held fixed). The conclusions
described in the literature above do not apply directly to our
results because they are based on a different equilibrium.

The primary mechanism that allows the break-up of the frozen-in
condition and allows reconnection to proceed is analyzed in
detail. The focus is on the presence of a strong guide field and
its effect on the breaking of the frozen-in constraint. The guide
field influence on the electron and ion motion is studied. The
mechanism which permits fast reconnection in the presence of a
guide field is studied; in particular, the simulations yield
several new results. First, the dynamics of fast reconnection in
the presence of a large guide field and high mass ratio is
explored, resulting in the scaling of the reconnection rate with
both of these parameters. Second, the simulations provide results
on the break-up mechanism of the frozen-in condition in the
presence of a strong guide field, a crucial problem in the physics
of reconnection. Third, the influence of the guide field on the
in-plane and out-of-plane ion and electron velocities is analyzed.
Finally, the physics of fast reconnection is discussed, showing
for the first time in a fully kinetic simulation with a strong
guide field the typical electron density pattern related to the
KAW dynamics previously predicted by theoretical studies [{\it
Kleva et al.}, 1995] and shown by fluid simulations [{\it Rogers
et al.}, 2003].

The paper is organized as follows. Section II describes the
physical problem and the numerical approach. Section III presents
the results of the simulations, studies the mechanism for
reconnection, analyzes the motion of ions and electrons, and
examines the scaling of the reconnection rate with the mass ratio
and the guide field.

\section{Physical system and simulation approach}

A Harris current sheet is considered in the $(x,z)$ plane [{\it
Harris}, 1962], with an initial magnetic field given by
\begin{equation}
\mathbf{B}_{0}(z)=B_0 \tanh(z/\lambda) \mathbf{e}_x+ B_{y0}
\mathbf{e}_y
\end{equation}
and a plasma density given by
\begin{equation}
n_0(z)=n_0 \, \hbox{sech} ^2(z/ \lambda)+n_b
\end{equation}

The GEM physical parameters are used [{\it Birn et al.}, 2001].
The temperature ratio is $T_e/T_i=0.2$, the current sheet
thickness is $\lambda=0.5 d_{i}$, the background density is
$n_b=0.2 n_0$, and the ion drift velocity in the $y$ direction is
$V_{i0}=1.67V_A$, where $V_A$ is the Alfv\'en velocity defined
with the density $n_0$ and the field $B_0$, and
$V_{e0}/V_{i0}=-T_{e0}/T_{i0}$. The ion inertial length,
$d_i=c/\omega_{pi}$, is defined using the density $n_0$. The
standard GEM challenge is modified by introducing an initially
uniform guide field $B_y=B_{y0}$, which eliminates the line of
null magnetic field. Different mass ratios are used, ranging from
$m_i/m_e=25$ (standard GEM mass ratio) to the physical mass ratio
for hydrogen, $m_i/m_e=1836$. Following {\it Birn et al.} [2001],
the Harris equilibrium is modified by introducing an initial flux
perturbation in the form
\begin{equation}
A_y=-A_{y0} \cos (2 \pi x / L_x) \cos ( \pi z /L_z)
\end{equation}
with $A_{y0}=0.1 B_0 c / \omega_{pi}$.

The boundary conditions for the particles and fields are periodic
in the $x$ direction.  Conducting boundary conditions are imposed
for the fields at the $z$ boundaries while reflecting boundary
conditions are used for the particles.

The simulations shown in the present paper are performed using two
PIC codes. CELESTE3D, an implicit PIC code, solves the full set of
Maxwell-Vlasov equations using the implicit moment method [{\it
Brackbill and Forslund}, 1985; {\it Vu and Brackbill}, 1992; {\it
Ricci et al.}, 2002a]. Maxwell's equations are discretized
implicitly in time, as

\begin{equation}
\left\{
\begin{array}{l}
\nabla \cdot \mathbf{E}^{\theta}= 4 \pi n^{\theta} \\
\displaystyle{\nabla \times \mathbf{E}^{\theta}=-\frac{1}{c} \frac{\mathbf{B}^1-\mathbf{B}^0}{\Delta t}}\\
\nabla \cdot \mathbf{B}^1=0\\
\nabla \times \mathbf{B}^{\theta}= \displaystyle{\frac{1}{c}
\frac{\mathbf{E}^1-\mathbf{E}^0}{\Delta t}+ \frac{4 \pi}{c}
\mathbf{J}^{1/2}}
\end{array}
\right.
\end{equation}
where the superscript 1 and 0 denote the new and old time levels,
and $\theta \in [1/2,1]$. Newton's equations for each particle are
also discretized implicitly in time:
\begin{equation}
\left\{
\begin{array}{l}
\mathbf{x}_p^1=\mathbf{x}^0_p+\mathbf{u}_p^{1/2} \Delta t \\
\mathbf{u}_p^1=\mathbf{u}^0_p+ \displaystyle{\frac{q_s \Delta
t}{m_s}} \left[ \mathbf{E}^{\theta}(\mathbf{x}_p^{1/2}) +
\frac{\mathbf{u}_p^{1/2} \times \mathbf{B}^0}{c} \right]
\end{array}
\right.
\end{equation}

The implicit moment method allows more rapid simulations on ion
length and time scales than explicit methods, while retaining the
kinetic effects of both electrons and ions. The explicit
simulation must observe the time step limit, $\Delta t < 2 /
\omega_{pe}$, and the mesh spacing required to avoid the finite
grid instability, $\Delta x \leq 2 \lambda_{De}$, with
$\omega_{pe}$ and $\lambda_{De}$ denoting the electron frequency
and the electron Debye length. These two constraints are replaced
in implicit simulations by an accuracy condition, $v_{th,e} \Delta
t < \Delta x$, whose principal effect is to determine how well
energy is conserved. In general, the accuracy condition permits
much larger $\Delta x$ and $\Delta t$ than possible with explicit
methods.

The explicit plasma simulation code NPIC is based on a well-known
explicit electromagnetic algorithm [{\it Morse and Nielson}, 1971;
{\it Forslund}, 1985]. In this full-Maxwell approach, the fields
are advanced using the scalar and vector potentials. Working in
the Coulomb gauge, the scalar potential is computed directly from
Poisson's equation, while the vector potential is advanced in time
using either a simple explicit algorithm [{\it Morse and Nielson},
1971] or a semi-implicit method that permits the time step to
exceed the Courant limit [{\it Forslund}, 1985]. Intuitively, this
corresponds to an implicit treatment of light waves, while the
rest of the algorithm remains explicit and the electron plasma
frequency and cyclotron motion are fully resolved. In this
manuscript, all simulations at low mass ratio $m_i/m_e=25$ were
performed with the simple explicit version of the field solver,
while two of the simulations at higher mass ratio ($m_i/m_e=180$,
$B_{y0}=0,B_0$) were performed with the semi-implicit version of
the field solver with a time step approximately three times larger
than the Courant limit.  For the strong guide field high mass
ratio case ($m_i/m_e=180$, $B_{y0}=5 B_0$), the simple explicit
field solver was employed since the time constraint imposed by the
guide field on the particle mover is more limiting.
Extensive comparisons between the
two versions of the field solver have revealed no significant
differences. The particle trajectories within NPIC are advanced
using the leapfrog technique, and particle moments are accumulated
with area weighting.  To run on a parallel computer, the code is
written using domain decomposition with calls to the MPI library.

The implicit PIC method, and particularly the code CELESTE3D, has
been applied to many problems in plasma physics in one dimension
[{\it Lapenta and Brackbill}, 1994; {\it Lapenta}, 2002], in two
dimensions [{\it Quest et al.}, 1983; {\it Forslund et al.}, 1984;
{\it Dreher et al.}, 1996; {\it Lapenta and Brackbill}, 1997; {\it
Lapenta and Brackbill}, 2002; {\it Ricci et al.}, 2002a; {\it
Ricci et al.}, 2002b; {\it Lapenta et al.}, 2003], and in three
dimensions [{\it Lapenta and Brackbill}, 2000; {\it Lapenta et
al.}, 2003]. NPIC has been used to study the dynamics of thin
current layers [{\it Daughton}, 2002; {\it Daughton}, 2003]

Implicit PIC methods are particularly suitable when simulating
systems with higher mass ratios. The cost of an explicit
simulation on ion time and space scales varies with the ion to
electron mass ratio as $(m_i/m_e)^{(d+2)/2}$, where $d$ is the
number of spatial dimensions [{\it Pritchett}, 2000]. For example,
in two dimensions, a simulation of the GEM challenge with
$m_i/m_e=1836$ is more than $5000$ times as expensive as one with
$m_i/m_e=25$ if the explicit method is used. In contrast, the cost
of an implicit simulation scales as $(m_i/m_e)^{1/2}$, as the time
step can be kept constant with respect to the ion plasma frequency
and the ratio $\omega_{ci}/\omega_{pi}$ is scaled as
$(m_i/m_e)^{1/2}$, in order to maintain the same Harris sheet
equilibrium [{\it Ricci et al.}, 2002b]. (The ratio $v_{th,e}/c$
is kept constant while $v_A/c$ is decreased when the mass ratio is
increased). The explicit PIC method resolves all relevant scales
within the plasma, and with a massively parallel computer, high
mass ratio simulations are now quite feasible in two dimensions.
In the present work, the explicit simulations are run on a machine
using as many as 128 nodes.

The simulations have been performed by the two codes with
remarkably different simulation parameters. With $m_i/m_e=25$,
CELESTE3D uses a $N_x \times N_z=64 \times 64$ grid, with time
step $\omega_{pi} \Delta t=0.3$, and 25 particles per species per
cell, for a total of $2 \cdot 10^5$ computational particles. The
high mass ratio simulations are performed by CELESTE3D with the
same simulation parameters, except for the $m_i/m_e=1836$ case
where the time step is reduced to $\omega_{pi} \Delta t=0.1$. With
$m_i/m_e=25$, NPIC employs a $N_x \times N_z= 1024 \times 512$
grid, $104 \cdot 10^6$ computational particles and a time step
corresponding to $\omega_{pi} \Delta t= 0.029$ for $B_{y0}=0,B_0$,
and $\omega_{pi} \Delta t =0.014$ at $B_{y0}=5 B_0$. For the high
mass ratio cases $m_i/m_e=180$, the simulation parameters are $N_x
\times N_z = 2560 \times 1280$, $1 \cdot 10^9$ particles, and a
time step $\omega_{pi} \Delta t = 0.019$ for $B_{y0}=0,B_0$, and
$\omega_{pi} \Delta t = 0.0064$ at $B_{y0}=5 B_0$.

The reconnected fluxes for mass ratios $25$ and $180$ from the two
codes show a remarkable agreement. A detailed comparison of the
electron dynamics for $m_i/m_e=25$ has been performed, which shows
that the physical mechanisms revealed by the two codes agree,
although CELESTE3D does not resolve all the electron scales. In
fact, it is important to note that even at the mass ratio $25$,
the CELESTE3D grid does not resolve the electron scales; in
particular, the electron Debye length $\lambda_e$, the electron
skin depth $d_e=c/\omega_{pe}$, the electron gyroradius $\rho_e$,
and $\rho_s$ ($\rho_s^2=c^2 m_i T_e / e^2 B_{y0}^2$) are not
resolved.

Certainly, the accuracy of wave-particle interactions on the fast
time scale is reduced using a coarser grid and a bigger time step.
However, results of the simulations shown here and many previous
simulations show that kinetic electrons seem to contribute
correctly on the ion time scales: Generally speaking, kinetic
electrons contribute inertial effects, anisotropic pressure, and
electron thermal transport on the ion time scales that would
otherwise have to be modelled if fluid electron equations were
used [{\it Forslund and Brackbill}, 1982; {\it Brackbill et al.},
1984; {\it Vu and Brackbill}, 1993; {\it Lapenta and Brackbill},
1996; {\it Ricci et al.}, 2002b].

Convergence studies of the implicit PIC method have already been
presented in {\it Brackbill and Forslund} [1985] and {\it
Brackbill and Vu} [1991]. The non-linear evolution of the LHDI
predicted by CELESTE3D [{\it Lapenta and Brackbill}, 2002], has
been confirmed by explicit results [{\it Lapenta et al.}, 2003].
Regarding the GEM challenge, it has already been shown that
results with CELESTE3D are comparable in detail with those of
explicit simulations [{\it Pritchett}, 2000] for $m_i/m_e=25$ and
$B_{y0}=0$ [{\it Ricci et al.}, 2002a]. A convergence study in
presence of a guide field has been performed. In Tab. I, the
dependence of conservation of the total energy of the plasma on
the time step is studied for $B_{y0}=5B_0$ and $m_i/m_e=25$. For
reference, the energy conservation for the explicit method is also
shown.

\section{Simulation results}

A set of simulations is performed, using different mass ratios
($m_i/m_e=25$, $m_i/m_e=180$, and $m_i/m_e=1836$) and different
guide fields:  the standard GEM challenge with $B_{y0}=0$,
$B_{y0}= B_{0}$ and $B_{y0}=5 B_{0}$, corresponding to $\beta=
\infty$, $\beta=1.2$, and $\beta=0.048$, in the center of the
current sheet [$\beta=8 \pi n_0(T_i+T_e)/(B_{y0}^2)$].

In all cases, the typical evolution of the magnetic flux and the
out-of-plane current is similar to the picture of magnetic
reconnection in the absence of a guide field provided by the GEM
challenge project [e.g., {\it Ricci et al.}, 2002a]. In
particular, in the presence of a guide field reconnection still
occurs but it requires a longer time and saturates at a lower
level. The current is considerably more filamentary and peaks of
negative current appear which are not present in the standard GEM
challenge with no guide field.

In Fig. 1, the reconnection rates for both NPIC and CELESTE3D
simulations are shown. The reconnected flux is measured as the
flux difference, $\Delta \Psi$, between the X and the O points.
All the simulations show a similar evolution. After slow initial
growth, which lasts until $t\omega_{ci} \approx 10$ (or longer,
for higher guide fields), reconnection enters a fast phase that
persists until the saturation level is reached. During the fast
reconnection phase, both NPIC and CELESTE3D (when simulations with
enhanced spatial resolution are performed) show multiple small
scale islands, which merge at later time into a single island. In
general, the reconnection rate decreases as the guide field
increases for all values of the mass ratio. However, a fast
reconnection phase is always present. The saturation level also
decreases with the guide field, because the out-of-plane magnetic
field influences the plasma motion reducing its compressibility.
It is also shown that the reconnection rate depends weakly on the
mass ratio for all the guide fields considered. Simulations are
performed with $m_i/m_e=25, 180, 1836$ with the implicit PIC code
CELESTE3D and with $m_i/m_e=25, 180$ for the explicit PIC code
NPIC. For the reconnected flux, the results of the explicit and
implicit simulations agree remarkably well for both mass ratios.
In some cases, the fast reconnection phase starts earlier in
implicit simulations; nevertheless, both the reconnection rate and
the saturation level is similar in the two codes. The later start
is probably due to the reduced initial noise in the explicit
simulation because of much larger number of particles.

>From an energetic point of view, the reconnection process causes a
decrease of the total magnetic energy, because the $x$-component
of the magnetic field is destroyed by the reconnection process. In
particular our results show that, while the energy related to the
$y$-component of the magnetic field is almost constant (it
slightly increases in the $B_y=0,B_0$ case), the $B_x$ field
energy decreases, and the energy of the $z$-component of the
magnetic field, which is created during the reconnection process,
grows. The lost magnetic energy is transferred to the ions and
electrons in form of kinetic energy.

In the following section, the mechanism which leads to the
break-up of the frozen-in condition for electrons is analyzed, and
the general motion of ions and electrons depending on the guide
field is studied. Finally, the mechanism of fast reconnection is
studied, when the whistler dynamics are suppressed by the presence
of the guide field.

\subsection{Break of the frozen-in condition}

Ohm's law in collisionless plasmas states that the reconnection
electric field, which is proportional to the reconnection rate,
can be expressed as [e.g., {\it Pritchett}, 2001; {\it Ricci et
al.}, 2002b]
\begin{eqnarray}
E_{y,rec}=&-& \frac{1}{c} \left( v_{ze}B_x - v_{xe} B_z\right)
-\frac{1}{en_e} \left( \frac{\partial P_{xye}}{\partial x} +
\frac{\partial P_{yze}}{\partial z} \right) \nonumber \\
&-& \frac{m_e}{e} \left( \frac{\partial v_{ye}}{\partial t} +
v_{xe} \frac{\partial v_{ye}}{\partial x} + v_{ze} \frac{\partial
v_{ye}}{\partial z}\right)
\end{eqnarray}

At the X point, the magnetic field components $B_x$ and $B_z$
vanish and the only contributions to the reconnection electric
field are given by gradients of the off-diagonal terms of the
electron pressure or by the terms related to electron inertial
effects.

In the zero guide field case, Kuznetsova {\it et al.} [2000] show
that the electrons become demagnetized near the X point, at
distances comparable to the electron meandering lengths,
\begin{equation}
d_{xe}=\left[ \frac{c^2 m_e T_e}{e^2 ( \partial B_z/ \partial x)
^2} \right] ^{1/4},d_{ze}=\left[ \frac{c^2 m_e T_e}{e^2 ( \partial
B_x/ \partial z) ^2} \right] ^{1/4}
\end{equation}
and execute a bounce motion which results in the presence of
off-diagonal terms of the electron pressure tensor. In this case,
the off-diagonal terms of the electron pressure are most important
in breaking the frozen-in condition [{\it Pritchett}, 2001; {\it
Ricci et al.}, 2002b]. In the presence of a weak guide field
($B_{y0}=0.3 B_{0}$, $B_{y0}=0.8 B_0$), evidence has been given
that the electron pressure is still the mechanism that allows
reconnection [{\it Hesse et al.}, 2002; {\it Yin and Winske},
2003].

Figures 2 and 3 show the non-ideal part of out-of-plane electric
field [i.e., $E_y+(v_{ze}B_x-v_{xe}B_z)/c$, which is the
difference between the electric field $E_y$ and the ideal terms
${\bf v}_e \times {\bf B} / c$], and the contribution of the
electron pressure terms close to the X point during the
reconnection process. Simulations performed by NPIC (Fig. 2) and
CELESTE3D (Fig. 3) with the guide fields $B_{y0}=0$, $B_{y0}=B_0$,
and $B_{y0}=5B_0$ are considered for a mass ratio $m_i/m_e=25$.

In the $B_{y0}=0$ case, the electron pressure tensor is the
dominant contribution at the X point [e.g., {\it Pritchett}, 2001;
{\it Ricci et al.}, 2002b] on a scale length of the order of the
electron meandering length [{\it Ricci et al.}, 2002b]. NPIC and
CELESTE3D results agree well, although NPIC results are more
refined because of the higher number of particles.

For $B_{y0}=B_0$ and $B_{y0}=5B_0$, the results of the two codes
look less alike. Both NPIC and CELESTE3D do indicate that the
contribution of the electron pressure tensor is responsible for
the break-up of the frozen-in condition for all the guide fields.
Moreover, the numerical value of the non-ideal electric field at
the X point and its pressure contribution, which is equal to the
reconnection rate, are similar in the explicit and implicit
simulations. However, the higher resolution of NPIC reveals that
the characteristic thickness of the break-up region is of the
order of $\rho_e$ in presence of a guide field. Nevertheless,
CELESTE3D appears to capture the break-up region for the
electrons.

The out-of-plane electron velocity evaluated from both simulations
shows that electron inertia alone cannot be responsible for the
break-up mechanism [Hesse {\it et al.}, 1999]. Thus, only the
pressure terms can be relevant at the X point, and the implicit
moment method spreads out the electron diffusion region from the
electron gyroradius to a scale linked to the grid spacing.

As an aside, it should be noted that in the presence of a guide
field the symmetry of the break-up region seen in the $B_{y0}=0$
case is lost and is replaced by a more complex non-symmetric
structure.

In Fig. 4, which shows results for the case $m_i/m_e=25$ and
$B_{y0}=5$ from NPIC, allows to study the contributions to the
non-ideal electric field (Fig. 4a). In Fig. 4b are shown the
off-diagonal terms of the electron pressure tensor, in Fig. 4c the
convective inertial terms, $v_{xe} \partial v_{ye} / \partial x +
v_{ze} \partial v_{ye} / \partial z$, and in Fig. 4d the inertial
term, computed as the difference between the non-ideal electric
field and the pressure contribution. As in the zero guide field
case, the pressure terms dominate in the region closest to the X
point, while the inertia terms are relevant at intermediate
lengths. The contribution of the term $\partial v_{ye} / \partial
t$ appears small because Fig. 4c and Fig. 4d are similar. The
ideal terms give the main contribution far away from the
reconnection region. According to these results, the importance of
the pressure terms does not decrease in presence of a guide field.
However, the spatial thickness of the region in which these terms
dominate is of order $\rho_e$ and thus decreases with increasing
guide field.

Figures 2-4 confirms the finding by Hesse {\it et al.} [2002] that
the electron pressure is the dominant non-ideal term with $B_{y0}
\approx B_0$, and extends it also to a plasma with lower $\beta$.
As an aside, it should be pointed out that simulations with
$B_{y0}=5 B_0$ that explore the electron break-up region are
computationally very expensive, as they require the accurate
resolution of the electron Larmor motion.

According to Hesse {\it et al.} [2002], the off-diagonal pressure
is generated by gradients of the electron flow velocity in the $y$
direction and by differences in the diagonal terms of the pressure
tensor, as [{\it Hesse et al.}, 2002]
\begin{equation}
P_{xye}=-\frac{P_{zze}}{\omega_{ce}}\frac{\partial v_{ye}}{\partial z} + \frac{B_x}{B_y} (P_{yye}-P_{zze})\\
\end{equation}
\label{eq1}
\begin{equation}
P_{yze}=-\frac{P_{xxe}}{\omega_{ce}}\frac{\partial v_{ye}}{\partial x} + \frac{B_z}{B_y} (P_{yye}-P_{xxe})\\
\label{eq2}
\end{equation}
where the heat flux has been ignored. It is assumed that $B_y \gg
B_x$ and $B_y \gg B_z$, that the diagonal components of the
pressure tensor are much larger than the off-diagonal components,
and that $\tau \ll v_e/L$, where $\tau$ is a typical evolution
scale, $v_e$ is a typical electron velocity, and $L$ is a typical
scale length.

Hesse {\it et al.} [2002] show a good match between the
anisotropies, estimated from Eqs. (8) and (9), and the values
obtained directly from the simulations, for the $B_{y0}= 0.8 B_0$
case. For the $B_{y0}=5B_0$ case, Fig. 5 compares the actual value
of $P_{xye}$ and $P_{yze}$ obtained from the simulation, with the
value computed from Eqs. (8) and (9), for the NPIC simulation.
Once again, there is a good agreement between the simulation
results and the theoretical predictions. The contribution of
$P_{zze}/ \omega_{ce}
\partial v_{ye}/\partial z$ is relevant in the evaluation of
$P_{xye}$, while the contribution of $P_{xxe}/ \omega_{ce}
\partial v_{ye}/\partial x$ is unimportant to $P_{yze}$. CELESTE3D
confirms these findings. The mechanism through which the
differences among the diagonal terms of the electron pressure
arises has been studied by Ricci {\it et al.} [2003] and can be
extended to higher guide field.

The agreement between our results and Hesse {\it et al.} [2002] is
two-fold. First, our simulations agree with the theoretical model
of Eqs. (8) and (9). Second, our results obtained with NPIC and
the implicit code CELESTE3D agree with the conclusions obtained by
Hesse {\it et al.}'s code for $B_{y0} \approx B_0$. Furthermore,
the conclusions by Hesse {\it et al.} [2002] are extended to
larger guide fields, a parameter regime not yet explored.

\subsection{Ion and electron motion}

When $B_{y0}=0$, the ions and electrons $\mathbf{E} \times
\mathbf{B}$ drift towards the X point along the $z$ direction (see
e.g. Pritchett [2001]). The ions become demagnetized at distances
of the order of a few $d_i$, because of the Hall effect, are
accelerated along the $y$ direction by the reconnection electric
field, $E_y$, and flow outwards in the $x$ direction at the
Alfv\'en speed, where they are diverted by the $B_z$ magnetic
field. The electrons follow a similar flow pattern, except that
they are demagnetized at shorter distances, of the order of the
electron meandering lengths [see Eq. (7)], and are expelled at
super-Alfv\'enic velocities. The whole ion and electron motion is
up-down and left-right symmetric.

The presence of a guide field rotates the $\mathbf{E} \times
\mathbf{B}$ motion, causes ions and electrons to drift in
directions not otherwise possible [{\it Yin and Winske}, 2003],
and destroys the symmetry with respect to the $z=0$ axes.  In
Figs. 6-7, the ion and electron motion in the $(x,z)$ plane is
represented in the presence of a guide field.

In all cases, the ions are diverted when they approach the X point
in an antisymmetric way with respect to the $x=0$ line. Their
outflow motion is primarily along $x$. The outflow region becomes
smaller as the guide field increases. The electron dynamics are
completely different and depend strongly on the guide field. In
the $B_{y0}=B_0$ case (Fig. 6) electrons exhibit a strong flow
along the separatrix. The motion is inward in the first and third
quadrant, and outward in the second and fourth quadrants. Our
simulation confirms the asymmetric motion of the electrons, which
has been shown theoretically to have an important role in the
reconnection process [{\it Kleva et al.}, 1995; {\it Biskamp},
1997; {\it Rogers et al.}, 2001; {\it Yin and Winske}, 2003]. In
the presence of a stronger guide field, $B_{y0}=5B_0$, (Fig. 7),
the electrons flow with a similar pattern to the $B_{y0}=B_0$
case, but the in-plane motion is more localized.

The electron motion along the $y$ direction (i.e., the
out-of-plane direction of the guide field) is also affected by the
guide field [{\it Horiuchi and Sato}, 1997]. This is shown in Fig.
8, where CELESTE3D and NPIC results are compared. When $B_{y0}=0$,
the ions are accelerated by the reconnection electric field,
$E_y$, at the X point along the $y$ direction. However, the $B_z$
field diverts the electrons, decreasing the $y$ velocity, and
forcing the outflow in the $x$ direction. In the presence of a
guide field, even far from the X point, the electrons maintain a
significant velocity in the $y$ direction, as they flow along the
magnetic field. The electron motion is concentrated at the
separatrix and the $y$ velocity increases with the guide field.
CELESTE3D results [{\it Ricci et al}, 2003] reveal also a
dependence of the out-of-plane velocity on the mass ratio, showing
that lighter electrons reach higher velocities. This particle
acceleration may have important consequences in active galactic
nuclei, extragalactic jets, solar flares and auroral arcs [{\it
Larrabee et al.}, 2003; {\it Nodes et al.}, 2003 and references
therein].

\subsection{Fast reconnection mechanism}

When $B_{y0}=0$, the off-diagonal terms of the electron pressure
provide the primary mechanism by which the electrons break the
frozen-in condition, and the Hall term in Ohm's law decouples
electron and ion motion and strongly enhances the reconnection
rate. Because of the Hall effect, the electron and ion motion
decouple at a distance of the order of $d_i$ and the whistler
dynamics are enabled. The whistler waves have a quadratic
dispersion relation ($\omega \propto k^2$) [{\it Biskamp}, 1997],
which allows fast reconnection, even when the diffusion region is
small [{\it Shay et al.}, 2001]. The typical signature of the Hall
effect is the presence of a quadrupolar out-of-plane magnetic
field [{\it Sonnerup}, 1979; {\it Terasawa}, 1983] which has also
been observed by some satellite observations [{\it \O ieroset et
al.}, 2001; {\it Mozer et al.}, 2002].

At low $\beta$, provided that $\beta > m_e/m_i$ (i.e., $\rho_s >
d_e$, $\rho_s^2=c^2 m_i (T_e+T_i) / e^2 B_{y0}^2$), it has been
pointed out that the whistler dynamics are pushed to smaller
scales [{\it Rogers et al.}, 2001], because of magnetic field
compression, $B_{y0} \nabla \cdot \mathbf{v}$, which remains
finite even if the motion is almost incompressible [{\it Biskamp},
1997]. Nevertheless, due to the electron pressure term, and in
particular the parallel gradient of the electron density,
$\nabla_\parallel n_e$, KAW dynamics arise. The KAWs are
characterized by a quadrupole density structure with a scale
length $\rho_s$, which replaces $d_i$ as the spatial scale of
interest in presence of a guide field [{\it Kleva et al.}, 1995;
{\it Biskamp}, 1997; {\it Rogers et al.}, 2001]. KAWs have the
same dispersion properties as whistler waves ($\omega \propto
k^2$) and enable fast reconnection [{\it Biskamp}, 1997]. At still
lower $\beta$ ($\beta < m_e/m_i$), ions and electrons are tightly
coupled, ions are forced to follow the electron dynamics, and fast
reconnection is not possible [{\it Biskamp et al.}, 1997; {\it
Ottaviani and Porcelli}, 1993]. All of our simulations have a
plasma $\beta$ that permits fast reconnection. The mechanism for
fast reconnection operative at various $\beta$ have been explored
with fluid model [{\it Rogers et al.}, 2003]. Here we present the
first systematic kinetic study of fast reconnection mechanism as a
function of the plasma $\beta$.

In Fig. 9, the out-of-plane magnetic field during the reconnection
process is plotted for different mass ratios and different guide
fields. In the zero guide field case, the out-of-plane magnetic
field presents the typical quadrupolar symmetric structure due to
the Hall effect [{\it Sonnerup}, 1979; {\it Terasawa}, 1983]. When
a guide field is added to the initial Harris sheet equilibrium,
the out-of-plane magnetic field is dramatically altered. The
quadrupolar structure due to the Hall effect, is weakened and
tilted at $B_{y0}=B_0$, and is unidentifiable for $B_{y0}=5B_0$.
Even if the pattern of the magnetic field does not depend on the
mass ratios, the maximum and minimum values are affected. The
reason for this is the out-of-plane magnetic field depends on the
in-plane current which is due to the decoupling of ion and
electron motion and the electron motion depends on the mass ratio,
influencing the out-of-plane magnetic field.

The width of the ion outflow region is shown in Fig. 10 for three
different guide fields, by examining the $x$-component of the ion
velocity, $v_{xi}$. It is remarkable that for all guide fields,
the ion outflow pattern is not influenced by the mass ratios. It
follows that, at least for the range of guide fields studied,
there is a mechanism that decouples the ion and electron dynamics
(the electron dynamics depend on the mass ratio). Without a guide
field, the outflow region is of the order of a few $d_i$. In the
presence of the guide field $B_{y0}=B_0$, the outflow width
decreases, and the outflow region with $B_{y0}=5 B_0$ is narrower
than with $B_{y0}=B_0$ case (Fig. 10). We note also that the scale
length of interest, $\rho_s$, of this regime decreases when the
guide field increases.

The electron density pattern in the presence of a guide field is
plotted in Fig. 11. The quadrupolar pattern close to the
reconnection region is predicted by theory [{\it Kleva et al.},
1995], and it is a distinctive feature of the fast reconnection
enabled by the KAW physics.

Table II summarizes the variation of the reconnection rates, with
the guide field and mass ratio. The average growth rates are
listed.

The reconnection rate decreases as the guide field increases and
the fast reconnection mechanism transitions from whistler dynamics
to KAW dynamics. Experimental results confirm this trend [{\it
Yamada et al.}, 1997] as well as previous numerical results [{\it
Pritchett}, 2001]. A scaling law for the reconnection rate has
been proposed with this same property [{\it Wang et al.}, 2000].
{\it Horiuchi and Sato} [1999] propose another scaling law for the
reconnection rate which shows a decrease of the reconnection rate
as the guide field increases and which applies to driven
reconnection.

The reconnection rate shows only a weak dependence on the mass
ratio. For $B_{y0}=0$, Shay and Drake [1998] have demonstrated
that the reconnection rate is insensitive to the physics that
breaks the frozen-in condition, as a consequence, is insensitive
to the electron mass. This is confirmed by previous kinetic
calculations [{\it Pritchett}, 2001] and extended to the physical
mass ratio by Ricci {\it et al.} [2002b].

\section{Conclusions}

By performing kinetic simulations of Harris current sheets with
different guide fields and different mass ratios, the physics of
magnetic reconnection in plasmas characterized by different
$\beta$ values has been studied.

A main result of these simulations is the scaling of the
reconnection rate with the guide field and the mass ratio, up to
physical values. As in the case of high $\beta$ plasmas, the
mechanism which breaks the electron frozen-in condition is
provided by the off-diagonal terms of the electron pressure
tensor. The simulations extend the results to high guide fields,
and demonstrate that the scale length of the diffusion region
passes form the electron meandering length for $B_{y0}=0$ to the
electron gyroradius in presence of a guide field. The simulations
indicate that the mechanism that allows fast reconnection changes
with $\beta$. For high $\beta$, the typical quadrupolar structure
of the out-of-plane magnetic field associated to the whistler
dynamics whistler dynamics is present in the simulations. This
mechanism allow the decoupling of electrons and ions. At low
$\beta$ (high guide fields), the KAW dynamics allows the
decoupling  and, the quadrupolar electron density pattern which
characterize the KAW and which had been predicted theoretically
and by fluid models before is revealed by the simulations. The
presence of a guide field causes additional components of the
$\mathbf{E} \times \mathbf{B}$ drift, which modify the ion and
electron motion causing asymmetric plasma flow.

The comparison between the implicit and the explicit codes has
shown a remarkable agreement for phenomena occurring on spatial
scales resolved by both codes (e.g., the reconnected flux, the
structure of the out-of-plane magnetic field, the electron
velocity). A phenomenon occurring on spatial scale not resolved by
CELESTE3D, like the mechanism of the electron break-up mechanism,
is still present in the implicit code, but its effect has been
spread out to a more extended spatial scale.

In closing, we note that an experimental setup has been built to
study experimentally the dependence of reconnection on the guide
field [{\it Furno et al.}, 2003] and we plan to compare our
simulation results with the experiments.

\begin{acknowledgments}
The authors gratefully acknowledge useful discussions with J.
Birn, J. Egedal, P. Gary, R.V.E. Lovelace, M. Ottaviani, F.
Porcelli, B.N. Rogers,  M. Yamada, and L. Yin. This research is
supported by the LDRD program at the Los Alamos National
Laboratory, by the United States Department of Energy, under
Contract No. W-7405-ENG-36 and by NASA, under the "Sun Earth
Connection Theory Program".
\end{acknowledgments}

%
%

\newpage

\begin{itemize}

\item

Fig. 1: The reconnected flux is plotted for $m_i/m_e=25$ and
$B_{y0}=0$ (a), $m_i/m_e=25$ and $B_{y0}=B_0$ (b), $m_i/m_e=25$
and $B_{y0}=5 B_0$ (c), $m_i/m_e=180$ and $B_{y0}=0$ (d),
$m_i/m_e=180$ and $B_{y0}=B_0$ (e), $m_i/m_e=180$ and $B_{y0}=5
B_0$ (f), $m_i/m_e=1836$ and $B_{y0}=0$ (g), $m_i/m_e=1836$ and
$B_{y0}=B_0$ (h), and $m_i/m_e=1836$ and $B_{y0}=5 B_0$ (i). The
reconnected flux is normalized to $B_0c/\omega_{pi}$. The results
from CELESTE3D (solid line) and NPIC (dashed) are plotted.

\item

Fig. 2: For $m_i/m_e=25$, results from NPIC are shown for the
non-ideal part of the reconnection electric field,
$E_y+(v_{ze}B_x-v_{xe}B_z)/c$, (a,c,e) and electron pressure
contribution to the electric field (b,d,f), $- 1 /e n_e (\partial
P_{xye} / \partial y + \partial P_{zye} /
\partial z)$. Both plots are color coded, and
normalized to $B_0 V_A /c$. The magnetic field lines are plotted
in all frames and the guide fields are $B_{y0}=0$ (a,b),
$B_{y0}=B_0$ (c,d), and $B_{y0}=5B_0$ (e,f). $E_y$ is normalized
to $B_0 v_A/c$. The results are plotted at a time when $\Delta
\Psi=1$. The data are averaged over 100 time slices separated by
10 time intervals with $\Delta t \omega_{pi}=0.014$.

\item

Fig. 3: The corresponding results to Fig. 2 from CELESTE3D are
shown. The data is averaged over a moving window of 100 time
steps, with $\Delta t \omega_{pi}=0.03$.

\item

Fig. 4: Contributions to the non-ideal reconnection electric field
$E_y+(v_{ze}B_x-v_{xe}B_z)/c$ (normalized to $B_0 V_A / c$) (a)
given by electron pressure terms, $(\partial P_{xye} / \partial y+
\partial P_{zye} / \partial z)$ (b); $v_{xe} \partial v_{ye} / \partial x + v_{ze}
\partial v_{ye} / \partial z$ (c), and by all the inertia terms,
evaluated as the difference between the non-ideal electric field
and the pressure contribution. We consider $m_i/m_e=25$,
$B_{y0}=5B_0$. The results are plotted at a time when $\Delta
\Psi=1$. These results are from NPIC, and represent average over
100 time slices separated each other by a time step $\Delta t
\omega_{pi}=0.14$.

\item

Fig. 5: The actual off-diagonal terms of the electron pressure
tensor from the NPIC simulation with $m_i/m_e=25$ and
$B_{y0}=5B_0$ are plotted. Shown are $P_{xye}$ (a), and $P_{zye}$
(c), and their values computed from Eqs. (8) (b), and (9) (d). The
results are plotted at a time when $\Delta \Psi=1$. The data is
averaged over 100 time slices separated by a time step $\Delta t
\omega_{pi}=0.14$.

\item

Fig. 6: Ion (a) and electron (b) motion in the $(x,z)$ plane is
shown for $m_i/m_e=25$ and $B_{y0}=B_0$. The results are plotted
at a time when $\Delta \Psi=1$. These results are from CELESTE3D.

\item

Fig. 7: Ion (a) and electron (b) motion in the $(x,z)$ plane is
shown for $m_i/m_e=25$ and $B_{y0}=5 B_0$. The results are plotted
at a time when $\Delta \Psi=1$. These results are from CELESTE3D.

\item

Fig. 8: The electron velocity, $v_{ye}$, is shown at a time when
$\Delta \Psi=1$, for $B_{y0}=0$ (a,b), $B_{y0}=B_0$ (c,d),
$B_{y0}=5B_0$ (e,f), and mass ratio $m_i/m_e=25$.  These results
are from CELESTE3D (a,c,e) and NPIC (b,d,f).

\item

Fig. 9: The magnetic field, $B_y$, is shown when $\Delta \Psi=1$,
for $B_{y0}=0$ (a,b), $B_{y0}=B_0$ (c,d), $B_{y0}=5B_0$ (e,f), and
mass ratio $m_i/m_e=25$ (a,c,e), $m_i/m_e=1836$ (b,d,f). These
results are from CELESTE3D.

\item

Fig. 10: The ion velocity, $v_{xi}$, is plotted when $\Delta
\Psi=1$, for $B_{y0}=0$ (a,b), $B_{y0}=B_0$ (c,d), $B_{y0}=5B_0$
(e,f), and mass ratio $m_i/m_e=25$ (a,c,e), $m_i/m_e=1836$
(b,d,f). These results are from CELESTE3D.

\item

Fig. 11: The electron density, $n_e$, is plotted for $m_i/m_e=25$
and $B_{y0}=5B_0$ at time $t\omega_{ci}=30$ in the reconnection
region. The results are from CELESTE3D (a) and NPIC (b).

\end{itemize}

\newpage

\textbf{Table I.} Decrease of the error in the energy
conservation, $\Delta E (t)=[ E_{tot}(t)-E_{tot}(0)]/E_{tot}(0)$,
at time $t \omega_{ci}=40 $ when the time step is reduced. We
consider a set of simulations with $m_i/m_e=25$, $B_y/B_{y0}=5$,
$64 \times 64$ grid points, and 200 particles per cell.
\bigskip

\begin{tabular}
[c]{lcc} \hline\hline
$\Delta t \omega_{pi}$ & $\Delta E$\\
\hline
0.30 & 0.232\\
0.15 & 0.076\\
0.08 & 0.034\\
0.03 & 0.019\\
explicit (0.014) & 5.6 $10^{-4}$ \\
\hline\hline
\end{tabular}

\bigskip
\bigskip

\textbf{Table II.} Averaged reconnection rates, normalized to $B_0
v_A/c$, as a function of the guide field and the mass ratio.

\bigskip

\begin{tabular}
[c]{l|ccc} \hline\hline
& $m_i/m_e=25$ & $m_i/m_e=180$ & $m_i/m_e=1836$ \\
\hline
$B_{y0}/B_0$=0 & 0.179 & 0.190 & 0.206\\
$B_{y0}/B_0$=1 & 0.141 & 0.164 & 0.182\\
$B_{y0}/B_0$=5 & 0.086 & 0.098 & 0.113\\
\hline\hline
\end{tabular}

\end{article}

\begin{thebibliography}{}

\bibitem{Birn2001}
Birn, J., et al., Geospace Environment Modelling (GEM) magnetic
reconnection challenge, {\it J. Geophys. Res.}, {\it 106}, 3715,
2001.

\bibitem{Biskamp1970}
Biskamp, D., Magnetic reconnection in plasmas, Cambridge
University Press (Cambridge, New York, 2000).

\bibitem{Biskamp1997}
Biskamp, D., Collisional and collisionless magnetic reconnection,
{\it Phys. Plasmas}, {\it 4}, 1964, 1997.

\bibitem{Biskamp1997}
Biskamp, D., E. Schwarz, J.F. Drake, Two-fluid theory of
collisionless magnetic reconnection, {\it Phys. Plasmas}, {\it 4},
1002, 1997.

\bibitem{Blackman1996}
Blackman, E.G., Reconnecting magnetic flux tubes as a source of in
situ acceleration in extragalactic radio sources, {\it Astrophys.
J. Lett.}, {\it 456}, LT87, 1996.

Bobrova, N.A., S.V. Bulanov, J.I. Sakai, and D. Sugiyama,
Force-free equilibria and reconnection of the magnetic field lines
in collisionless plasma configurations, {\it Phys. Plasmas}, {\it
8}, 759, 2001.

\bibitem{Brackbill1984}
Brackbill, J. U., D. W. Forslund, K. B. Quest, and D. Winske,
Nonlinear evolution of the lower-hybrid drift instability, {\it
Phys. Fluids}, {\it 27}, 2682, 1984.

\bibitem{Brackbill1985}
Brackbill, J. U. and D. W. Forslund, Simulation of low frequency,
electromagnetic phenomena in plasmas, in {\it Multiple Times
Scales}, J.U. Brackbill and B.I. Cohen Eds., (Accademic Press,
Orlando, 1985), pp. 271-310.

\bibitem{Daughton2002}
Daughton, W., Nonlinear dynamics of thin current sheets, {\it
Phys. Plasmas}, {\it 9}, 3668, 2002.

\bibitem{Daughton2003}
Daughton, W., Electromagnetic properties of the lower-hybrid drift
instability in a thin current sheet, {\it Phys. Plasmas}, {\it
10}, 3103, 2003.

\bibitem{Drake2002}
Drake, J.F., M. Swisdak, C. Cattell, M.A. Shay, B.N. Rogers, A.
Zeiler, Formation of electron holes and particle energization
during magnetic reconnection, {\it Science}, {\it 299}, 873
(2003).

\bibitem{Dreher}
Dreher, J., U. Arendt, and K. Schindler, Particle simulations in
collisionless reconnection in magnetotail configuration including
electron dynamics, {\it J. Geophys. Res.}, {\it 101}, 27375
(1996).

\bibitem{Forslund}
Forslund, D. W. and J. U. Brackbill, Magnetic-field induced
surface transport on laser-irradiated foils, {\it Phys. Rev.
Lett.}, {\it 48}, 1614, 1982.

\bibitem{Forslund1984}
Forslund, D.W., K.B. Quest, J.U. Brackbill, and K. Lee,
Collisionless dissipation in quasi-perpendicular shocks, {\it J.
Geophys. Res.}, {\it 89}, 2142, 1984.

\bibitem{Forslund1985}
Forslund, D.,  Fundamentals of Plasma Simulation, {\it Space
Science Reviews}, {\it 42}, 3, 1985.

\bibitem{Furno}
Furno, I., T. Intrator, E. Torbert, C. Carey, M. D. Cash, J. K.
Campbell, W. J. Fienup, C. A. Werley, G. A. Wurden, and G. Fiksel,
Reconnection Scaling Experiment: a new device for three
dimensional magnetic reconnection studies, {\it Rev. Sci.
Instrum.}, {\it 74}, 2324, 2003.

\bibitem{Gekelman1991}
Gekelman, W., H. Pfister, Z. Lucky, J. Bamber, D. Leneman, and J.
Maggs, Design, construction, and properties of the large plasma
research device - the LAPD at UCLA, {\it Rev. Sci. Instrum.}, {\it
62}, 2875, 1991.

\bibitem{Harris1962}
Harris, E.G., On a plasma sheath separating regions of oppositely
directed fiels, {\it Nuovo Cimento Soc. Ital. fis. A-D}, {\it 23},
115, 1962.

\bibitem{Hesse1999}
Hesse, M., K. Schindler, J. Birn, and M. Kuznetsova, {\it The
diffusion region in collisionless magnetic reconnection}, {\it
Phys. Plasmas}, {\it 6}, 1781, 1999.

\bibitem{Hesse2002}
Hesse, M., M. Kuznetsova, and M. Hoshino, The structure of the
dissipation region for component reconnection: Particle
simulations, {\it Geophys. Res. Lett.}, {\it 29}, 2001GL014714,
2002.

\bibitem{Horiuchi1999}
Horiuchi, R., and T. Sato, Particle simulation study of
collisionless driven reconnection in sheared magnetic field, {\it
Phys. Plasmas}, {\it 4}, 277, 1997.

\bibitem{Hoshino1983}
Hoshino, M., and A. Nishida, Numerical simulation of the dayside
reconnection, {\it J. Geophys. Res.}, {\it 88}, 6926, 1983.

\bibitem{Hoshino1987}
Hoshino, M., Electrostatic effect for collisionless tearing mode,
{\it J. Geophys. Res.}, {\it 92}, 7368, 1987.

\bibitem{Kleva1995}
Kleva, R.G., J.F. Drake, and F.L. Waelbroeck, Fast reconnection in
high temperature plasmas, {\it Phys. Plasmas}, {\it 2}, 23, 1995.

\bibitem{Kuznetsova2000}
Kuznetsova, M.M., M. Hesse, and D. Winske, Toward a transport
model of collisionless magnetic reconnection, {\it J. Geophys.
Res.}, {\it 105}, 7601, 2000.

\bibitem{Lapenta1994}
Lapenta, G., and J.U. Brackbill, Dynamic and selective control of
the number of particles in kinetic plasma simulations, {\it J.
Comp. Phys.}, {\it 115}, 213, 1994.

\bibitem{Lapenta1996}
Lapenta, G., and J.U. Brackbill, Contact discontinuities in
collisionless plasmas: a comparison of hybrid and kinetic
simulation, {\it Geophys. Res. Lett.}, {\it 23}, 1713, 1996.

\bibitem{Lapenta1997}
Lapenta, G., and J.U. Brackbill, A kinetic theory for the
drift-kink instability, {\it J. Geophys. Res.}, {\it 102}, 27099,
1997.

\bibitem{Lapenta2000}
Lapenta, G., and J. U. Brackbill, 3D reconnection due to oblique
modes: a simulation of Harris current sheets, {\it Nonlinear
Processes Geophys.}, {\it 7}, 151, 2000.

\bibitem{Lapent}
Lapenta, G., Particle rezoning for multidimensional kinetic
particle-in-cell simulations, {\it J. Comp. Phys.}, {\it 181},
317, 2002.

\bibitem{Lapenta2002}
Lapenta, G., and J.U. Brackbill, Nonlinear evolution of the lower
hybrid drift instability: Current sheet thinning and kinking, {\it
Phys. Plasmas}, {\it 9}, 1544, 2002.

\bibitem{Lapenta2003}
Lapenta, G., J.U. Brackbill, and W. Daughton, The unexpected role
of the lower hybrid drift instability in magnetic reconnection in
three dimensions, {\it Phys. Plasmas}, {\it 10}, 1577, 2003.

\bibitem{Larrabee2003}
Larrabee, D.A., R.V.E. Lovelace, and M.M. Romanova, Lepton
acceleration by relativistic collisionless magnetic reconnection,
{\it Astrophys. J.}, {\it 586}, 72, 2003.

\bibitem{Lesch1997}
Lesch, H. and G.T. Birk, Particle acceleration by magnetic
field-aligned electric fields in active galactic nuclei, {\it
Astron. Astrophys.}, {\it 324}, 461, 1997.

\bibitem{Morse1971}
Morse, R., and C. Nielson, Numerical simulation of Weibel
instability in one and two dimensions, Phys. Fluids, {\it 14},
830, 1971.

\bibitem{Mozer}
Mozer, F.S., S.D. Bale, and T.D. Phan, Evidence of diffusion
regions at the subsolar magnetopause crossing, {\it Phys. Rev.
Lett.}, {\it 89}, 015002, 2002.

\bibitem{Nishida1978}
Nishidam, A., Geomagnetic Diagnostics of the Magnetosphere
(Springer-Verlag, New York, 1978).

\bibitem{Nishimura2003}
Nishimura, K., S.P. Gary, H. Li, S.A. Colgate, Magnetic
reconnection in a force-free plasma: Simulations of micro- and
macro instabilities, {\it Phys. Plasmas}, {\it 10}, 347 (2002).

\bibitem{Nodes2003}
Nodes C., G.T. Birk, H. Lesch, and R. Schopper, Particle
acceleration in three-dimensional tearing configuration, {\it
Phys. Plasmas}, {\it 10}, 835, 2003.

\bibitem{Oieroset2001}
\O ieroset, M., T.D. Phan, M. Fujimoto, R.P. Lin, and R.P.
Lepping, In situ detection of collisionless reconnection in the
Earth's Magnetotail, {\it Nature}, {\it 412}, 414, 2001.

\bibitem{Ottaviani1993}
Ottaviani, M., and F. Porcelli, Nonlinear collisionless magnetic
reconnection, {\it Phys. Rev. Lett.}, {\it 71}, 3802, 1993.

\bibitem{Priest1982}
Priest, E.R., Solar Magnetohydrodynamics, (Reidel, Dordrecht,
1982).

\bibitem{Priest2000}
Priest, E.R. and T. Forbes, Magnetic Reconnection: MHD Theory and
Applications (Cambridge University Press, Cambridge, England,
2000).

\bibitem{Pritchett2000}
Pritchett, P. L., Particle-in-Cell simulations of magnetosphere
electrodynamics, {\it IEEE Trans. Plasma Sci.}, {\it 28}, 1976,
2000.

\bibitem{Pritchett2001a}
Pritchett, P.L., Geospace Environment Modelling magnetic
reconnection challenge: Simulation with a full particle
electromagnetic code, {\it J. Geophys. Res.}, {\it 106}, 3783,
2001.

\bibitem{Quest1985}
Quest, K.B., D.W. Forslund, J.U. Brackbill, and K. Lee,
Collisionless dissipation processes in quasi-parallel shocks, {\it
Geophys. Res. Lett.}, {\it 10}, 471 (1983).

\bibitem{Ricci2002a}
Ricci, P.,  G. Lapenta, and J.U. Brackbill, A simplified implicit
Maxwell solver, {\it J. Comp. Phys.}, {\it 183}, 117, 2002a.

\bibitem{Ricci2002b}
Ricci, P.,  G. Lapenta, and J.U. Brackbill, GEM reconnection
challenge: implicit kinetic simulations with the physical mass
ratio, {\it Geophys. Res. Lett.}, {\it 29}(23), 2008,
10.1029/2002GL015314, 2002b.

\bibitem{Ricci2003}
Ricci, P.,  G. Lapenta, and J.U. Brackbill, Electron acceleration
and heating in collisionless magnetic reconnection, {\it Phys.
Plasmas}, {\it 10}, 3554, 2003.

\bibitem{Rogers2001}
Rogers, B.N., R.E. Denton, J.F. Drake, and M.A. Shay, Role of
dispersive waves in collisonless magnetic reconnection, {\it Phys.
Rev. Lett.}, {\it 87}, 195004, 2001.

\bibitem{Rogers2003}
Rogers, B.N., R.E. Denton, and J.F. Drake, Signatures of
collisionless magnetic reconnection, {\it J. Geophys. Res.}, {\it
108}, 1111, 2003.

\bibitem{Romanova1992}
Romanova, M.M., and R.V.E. Lovelace, Magnetic-field, reconnection,
and particle-acceleration in extragalatic jets, {\it Astron.
Astrophys.}, {\it 262}, 26, 1992.

\bibitem{Shay1998}
Shay, M.A., and J.F. Drake, The role of electron dissipation on
the rate of collisionless reconnection, {\it Geophys. Res. Lett.},
{\it 25}, 3759, 1998.

\bibitem{Shay2001}
Shay, M. A., J. F. Drake, B.N. Rogers, and R. E. Denton,
Alfv\'enic collisionless magnetic reconnection and the Hall term,
{\it J. Geophys. Res.}, {\it 106}, 3759, 2001.

\bibitem{Sonnerup}
Sonnerup, B.U.O., Magnetic Field Reconnection, in {\it Solar
System Plasma Physics}, vol. III, edited by L.T. Lanzerotti, C.F.
Kennel, and E.N. Parker, p. 45, North-Holland, New York, 1979.

\bibitem{Taylor}
Taylor, J.B., Relaxation and magnetic reconnection in plasmas,
{\it Rev. Mod. Phys.}, {\it 28}, 243, 1986.

\bibitem{Terasawa}
Terasawa, T., Hall current effect on tearing mode instability,
{\it Geophys. Res. Lett.}, {\it 10}, 475, 1983.

\bibitem{Vu1992}
Vu, H. X. and J. U. Brackbill, CELEST1D: An implicit, fully
kinetic model for low-frequecy Electromagnetic plasma simulation,
{\it Comput. Phys. Commun.}, {\it 69}, 253, 1992.

\bibitem{Vu1993}
Vu, H. X. and J. U. Brackbill, Electron kinetic effects in
switch-off slow shocks, {\it J. Geophys. Res.}, {\it 20}, 2015,
1993.

\bibitem{Wang2000}
Wang., X., A. Bhattacharjee, and Z.W. Ma, Collisionless
reconnection: effects of Hall current and electron pressure
gradient, {\it J. Geophys. Res.}, {\it 105}, 27633, 2000.

\bibitem{Yamada1999}
Yamada, M., Review of controlled laboratory experiments on physics
of magnetic reconnection, {\it J. Geophys. Res.}, {\it 104},
14529, 1999.

\bibitem{Yin2003}
Yin, L., and D. Winske, Plasma pressure tensor effects on
reconnection: hybrid and Hall-magnetohydrodynamics simulations,
{\it Phys. Plasmas}, {\it 10}, 1595, 2003.

%
%
\end{thebibliography}
\end{document}